\documentclass[showpacs,preprintnumbers,pre,twocolumn,nofootinbib]{revtex4}
\usepackage{amsmath}
\usepackage{graphicx}
\usepackage{dcolumn}
\usepackage{bm}

\input{tcilatex}

\begin{document}

\preprint{}
\title{Sensitive predators and endurant preys in an ecosystem driven by
correlated noises}
\author{Wei-Rong Zhong}
\author{Yuan-Zhi Shao}
\altaffiliation[ ]{Corresponding Author}
\email{stssyz@zsu.edu.cn}
\author{Zhen-Hui He}
\affiliation{Department of Physics, Sun Yat-sen University, 510275 Guangzhou, People's
Republic of China}

\begin{abstract}
We investigate the Volterra ecosystem driven by correlated noises. The
competition of the predators induces an increasing in population density of
the predators. The competition of the preys, however, leads the predators to
decay. The predators may have better stability under strong correlated
noises. The predators undergo a sensitivity to a random environment, whereas
the preys exhibit a surprising endurance to this stochasticity.
\end{abstract}

\maketitle

Predator-prey ecosystems are one of the most amazing models in interacting
population biology [1-3]. The prey-predator model and its derived ones are
applied in a wide range of fields, for example, tumors, measles, epidemics
and food chains [4-6]. Though predator-prey model suggests that the simple
interactions can result in periodic behavior of the population, it is still
an unrealistic assumption. To fill this deficiency, scientists struggle to
find more realistic approaches. Stochastic models are so far better
selections, including individual level model (ILM)[7] and two-species
stochastic population model (SPM) [8-10]. The anterior probes a situation
when the number of individuals is finite, and the later describes a strictly
infinite one.

Surprising phenomena emerge sometimes just because there exists correlation
or collaboration; one of the most well known example is the self-organized
behavior [11]. The population in ecosystems, including prey and predators,
are affected by simple stochastic processes of morality, reproduction, and
predation. Differential equations with stochastic components match their
realities better than the deterministic ones. Due to the difficulty and
complexity in mathematical analysis, previous researches in prey-predator
models only consider independent noises [8-10]. Regrettably, the correlated
noises, which are more realistic tentation to depict true nature than
independent ones, are easily left in the basket. We all know that stochastic
processes in ecosystems may come from the same origin like disaster, thus it
is reasonable to consider the correlation between noises. In this paper, we
focus on the two-species model with correlated stochastic components for the
first time. The correlated noises we consider here, which distinguish from
independent ones Cai and Lin have concerned with in their latest paper [8],
induce some novel phenomena in a prey-predator ecosystem that are not found
before.

We study two-species models to elucidate the mechanism of interesting
influence of correlated noises on the stochastic dynamical ecosystems.
Consider Volterra's model with prey resource limitation, described by the
stochastic differential equation 
\begin{eqnarray}
\frac{dX}{dt} &=&r_{1}X(1-\frac{X}{K})-k_{1}XY+X(1-\frac{X}{K})\xi (t), \\
\frac{dY}{dt} &=&-r_{2}Y+k_{2}XY+Y\eta (t),
\end{eqnarray}%
where $X\geq 0$ and $Y\geq 0$ denote the densities of individuals in a
population density of preys and predators, respectively. In the absence of
noise, they will evolve into a stable value $(X_{0},Y_{0})$ as $t\rightarrow
\infty .$ $r_{1},r_{2}$ are the corresponding reproduction rate. $K$ is the
capacity of prey population, that is, the population size for $t\rightarrow
\infty $ if $Y\equiv 0$. The coefficient $k_{1}$ quantifies the impact which
an individual predator has on the reproduction rate of an individual prey.
Conversely, $k_{2}$ gives the impact which an individual prey has on the
reproduction rate of a predator. $\xi (t),\eta (t)$ are correlated, Gaussian
white noises satisfying%
\begin{eqnarray}
\langle \xi (t)\xi (t^{\prime })\rangle &=&2M_{1}\delta (t-t^{\prime }), \\
\langle \eta (t)\eta (t^{\prime })\rangle &=&2M_{2}\delta (t-t^{\prime }), \\
\langle \xi (t)\eta (t^{\prime })\rangle &=&2\lambda \sqrt{M_{1}M_{2}}\delta
(t-t^{\prime }),
\end{eqnarray}%
in which $M_{1},M_{2}$ are the intensities of noises; $\lambda $, ranges
from zero to one, denotes the correlation coefficient between $\xi (t)$ and $%
\eta (t)$, $\delta (t-t^{\prime })$ is Dirac delta function under different
moments.

An approach to describing stochastic noise is using three important ideas:
samples, events, and probability [12]. For example, consider a large
population, say $N$, of preys. These are the samples. The events we consider
are the collection of all possible aggregations of these preys. The
probability of an event is simple $(1/N)$ times the number of samples in the
event. Therefore, our notation for the stationary probability distribution
(SPD) of an event $X$ (a population density of the preys) or $Y$ (a
population density of the predators) is%
\begin{eqnarray}
P(X) &=&\frac{Number\text{ }of\text{ }preys\text{ }in\text{ }X}{N_{1}}, \\
P(Y) &=&\frac{Number\text{ }of\text{ }predators\text{ }in\text{ }Y}{N_{2}},
\end{eqnarray}%
in which $N_{1},N_{2}$ are all preys and predators, respectively. Consider $%
P(X)$ and $P(Y)$ are independent, the joint stationary probability
distribution (JSPD) of $(X$,$Y)$ is $P(X,Y)=P(X)P(Y)$ satisfying $%
\int_{0}^{\infty }\int_{0}^{\infty }P(X,Y)dXdY=1.$

In general, the probability distribution is still a qualitative parameter,
though noise-induced transition can be depicted by observing its changing
trend. Here we define a quantitative parameter, i.e. the mean density of
population, which is useful to display the influence of noises on the
prey-predator ecosystems. The mean densities of the preys and predators are
respectively written as%
\begin{eqnarray}
\langle X\rangle &=&\int_{0}^{\infty }\int_{0}^{\infty }XP(X,Y)dXdY, \\
\langle Y\rangle &=&\int_{0}^{\infty }\int_{0}^{\infty }YP(X,Y)dXdY,
\end{eqnarray}%
and their variances are $\sigma _{X}^{2}=\langle X^{2}\rangle -\langle
X\rangle ^{2},$ and $\sigma _{Y}^{2}=\langle Y^{2}\rangle -\langle Y\rangle
^{2},$ respectively.

One of the approaches to solve Eq.(1) and (2) is deriving the equivalent
stationary Fokker-Planck equation [8], but that is possible only toward two
non-correlated noises. Once consider the correlation of noises, it is
difficult to derive and solve the Fokker-Planck equation. Here we apply a
real-time simulation of Eq.(1) and (2) to obtain the joint stationary
probability distributions of the preys and predators densities. JSPD is
useful to represent the influence of noises on the preys and predators. The
quantitative parameters are also calculated for deep analysis. For example,
the mean density of the predators is used to display their long-term
behavior.

Figure 1 shows the peak of the joint stationary probability distributions
drops and shifts to position $(0,0)$ with $M_{2}$ increasing. With an
increase in $M_{1},$ similar effects are observed, and they are not plotted
here for simplicity. Obviously, lower peak height and more large preys and
predators densities mean less system stability.%
\ifcase\msipdfoutput
\FRAME{ftbpFU}{3.4212in}{2.6446in}{0pt}{\Qcb{Joint stationary probability
distributions under different noise intensities. The parameters are $%
M_{1}=0.05,$ $\lambda =0.0,$ $(a)-(d)$ $M_{2}=0.1,$ $0.5,$ $2.0,$ $5.0$. }%
}{}{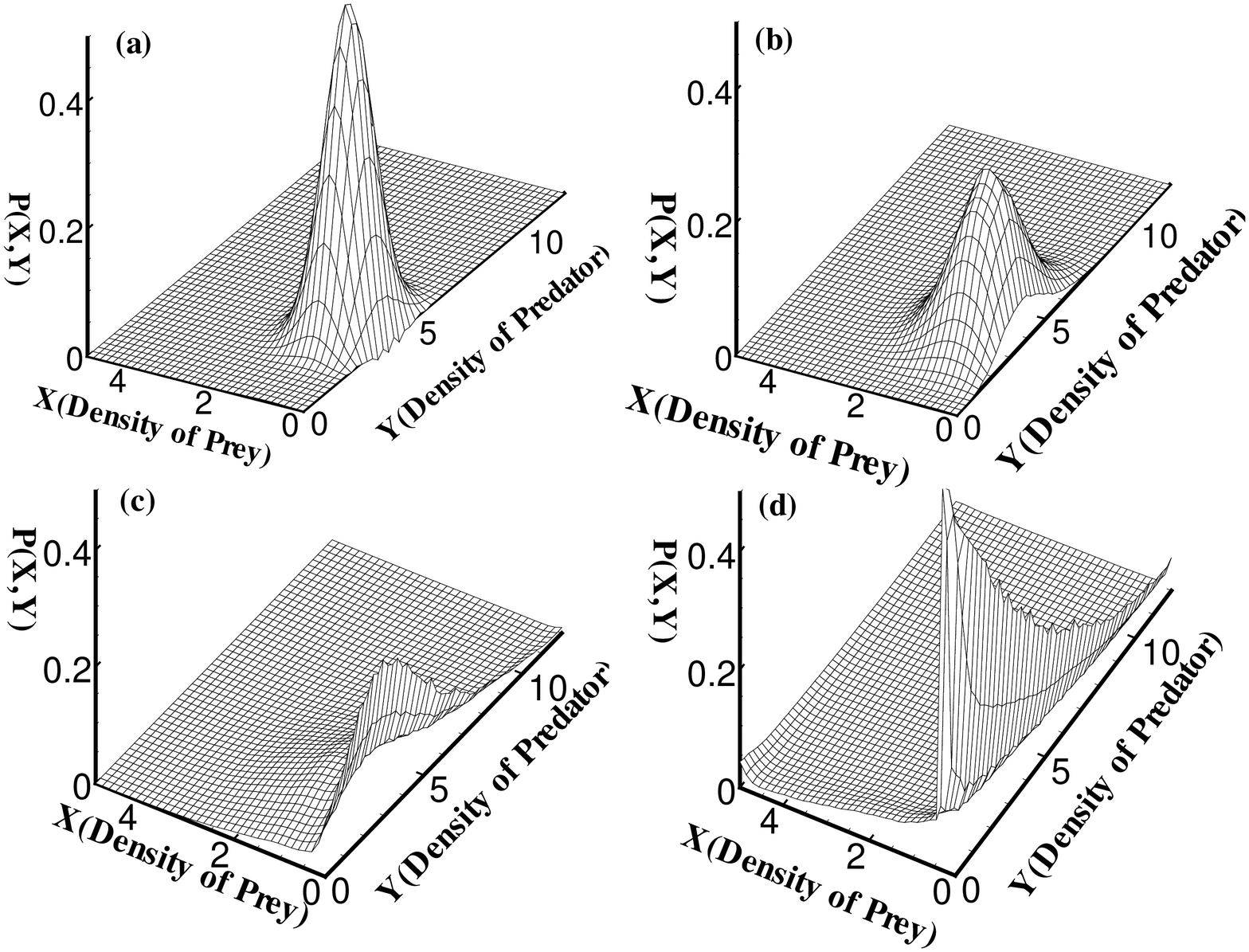}{%
\special{language "Scientific Word";type "GRAPHIC";maintain-aspect-ratio
TRUE;display "USEDEF";valid_file "F";width 3.4212in;height 2.6446in;depth
0pt;original-width 10.7972in;original-height 8.3281in;cropleft "0";croptop
"1";cropright "1";cropbottom "0";filename 'fig1.eps';file-properties
"XNPEU";}}%
\else
\begin{figure}[ptb]\begin{center}
\includegraphics[
natheight=8.3281in, natwidth=10.7972in, height=2.6446in, width=3.4212in]
{F:/Documents and Settings/Shao.ZSU-IPW2XA3O6HH/桌面/共享文件/Papers/Prey&Predator/graphics/fig1__1.pdf}%
\caption{Joint stationary probability distributions under different noise
intensities. The parameters are $M_{1}=0.05,$ $\protect\lambda =0.0,$ $%
(a)-(d)$ $M_{2}=0.1,$ $0.5,$ $2.0,$ $5.0$. }
\end{center}\end{figure}%
\fi

It is comprehensible that noises can induce un-stability. Consider a long
term, whether the noises will do good to the preys and predators or not is
still not known. So quantitative parameters, e.g. the mean densities of the
preys and predators are helpful to do that. Only according to the JSPD, it
looks as if there are no differences between the influence of noises on the
preys and those on the predators. Figure 2, however, illustrates their
distinctions. The average predator density increases with $M_{2}$, and
decreases with $M_{1}$. Obviously, $\eta (t)$ does good to the predators,
but $\xi (t)$ does not. Provided that the former is regarded as the
predators competition and the latter as the preys competition, it is not
difficult to comprehend why predators benefit from their competitions, and
why not from the prey competitions. Surprisingly, the preys densities
properly maintain their intactness in spite of the changing in $M_{1}$ or $%
M_{2}$. Why do they happen like that? Here we suggested an important reason,
which is easily understood from Eq.(1) and (2): both of the growth rate and
source limitation keep the prey's stability, whereas the only death rate can
not keeps the predator's stability. Perhaps this is why the predators are
more sensitive to the stochastic noises than the preys.%
\ifcase\msipdfoutput
\FRAME{ftbpFU}{3.3382in}{2.6446in}{0pt}{\Qcb{Relationship of mean densities
of the preys and predators with the noises intensities as $\lambda =0.0.$ $%
(a)$ $M_{1}=0.05,(b)$ $M_{2}=0.1.$}}{}{fig2.eps}{%
\special{language "Scientific Word";type "GRAPHIC";maintain-aspect-ratio
TRUE;display "USEDEF";valid_file "F";width 3.3382in;height 2.6446in;depth
0pt;original-width 4.401in;original-height 3.4783in;cropleft "0";croptop
"1";cropright "1";cropbottom "0";filename '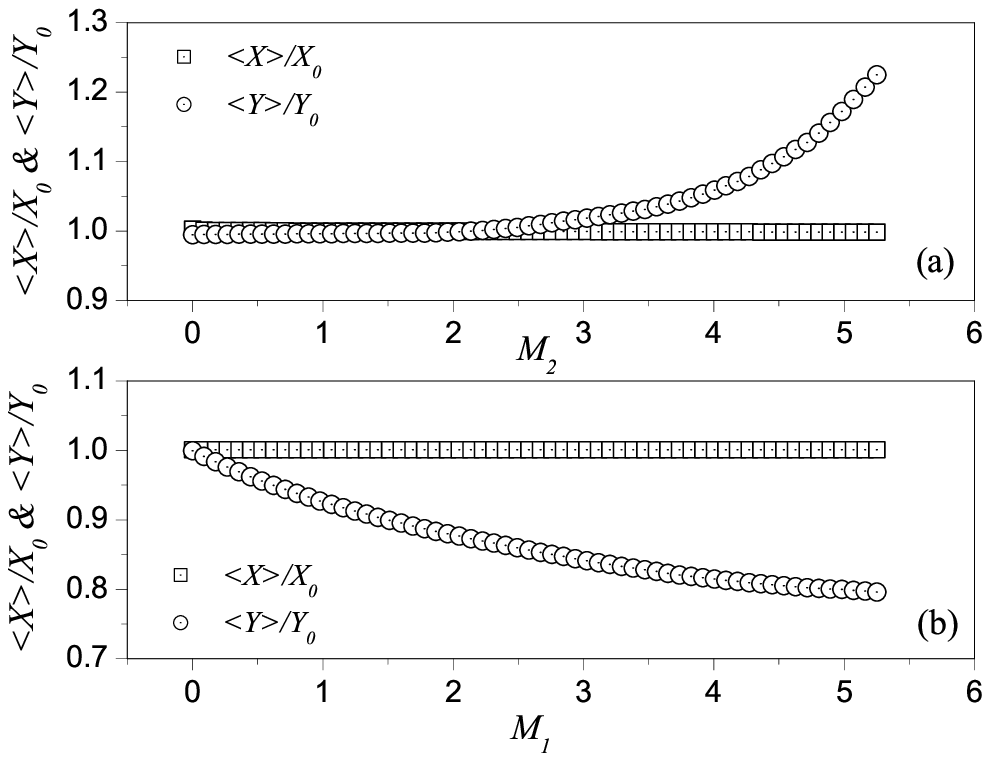';file-properties
"XNPEU";}}%
\else
\begin{figure}[ptb]\begin{center}
\includegraphics[
natheight=3.4783in, natwidth=4.401in, height=2.6446in, width=3.3382in]
{F:/Documents and Settings/Shao.ZSU-IPW2XA3O6HH/桌面/共享文件/Papers/Prey&Predator/graphics/fig2__2.pdf}%
\caption{Relationship of mean densities of the preys and predators with the
noises intensities as $\protect\lambda =0.0.$ $(a)$ $M_{1}=0.05,(b)$ $%
M_{2}=0.1.$}
\end{center}\end{figure}%
\fi

The counterpart distinctions induced by the correlation are shown in Fig.3
and Fig.4. Figure 3 displays that the stationary probability distributions
of the prey does not change with the noise correlation. The stationary
probability distributions have the same fitting curve under different
correlations. Their variances are also invariable (see the inset of Fig.3).
This indicates that the correlation has little effect on the preys. Unlike
what have happened on the preys, figure 4 shows that the peak height of the
stationary probability distributions of the predators increases with the
correlation (see the corresponding fitting curves), and its variance, shown
in the inset of Fig.4, decreases with the correlation. Like what we have
mentioned above, $\xi (t)$ and $\eta (t)$ refer to the preys and predators
competitions, respectively. We suggest that the correlation between noises
refers to the synchronous components of the competitions, and thus it is
possible for the predators to maintain more stable SPD provided that they
adjust their competition to those of the preys.%
\ifcase\msipdfoutput
\FRAME{ftbpFU}{3.4757in}{2.7588in}{0pt}{\Qcb{Stationary probability
distributions of the preys under different correlations. The parameters are $%
M_{1}=0.05,$ $M_{2}=0.1$. Inset: Variances of SPD against the correlations, $%
\lambda $.}}{}{fig3.eps}{%
\special{language "Scientific Word";type "GRAPHIC";maintain-aspect-ratio
TRUE;display "USEDEF";valid_file "F";width 3.4757in;height 2.7588in;depth
0pt;original-width 4.0932in;original-height 3.243in;cropleft "0";croptop
"1";cropright "1";cropbottom "0";filename '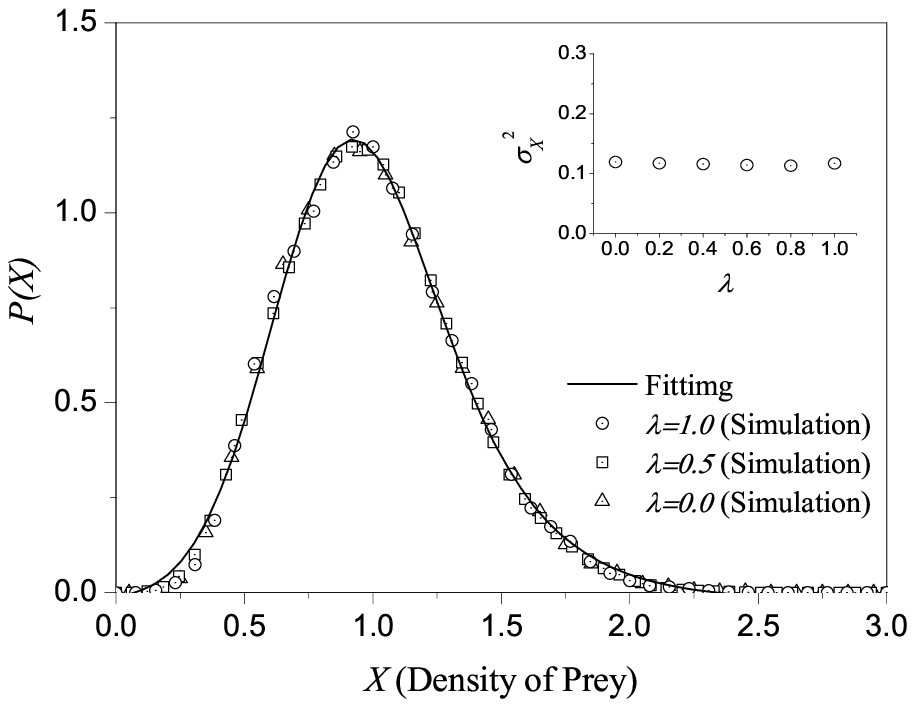';file-properties
"XNPEU";}}%
\else
\begin{figure}[ptb]\begin{center}
\includegraphics[
natheight=3.243in, natwidth=4.0932in, height=2.7588in, width=3.4757in]
{F:/Documents and Settings/Shao.ZSU-IPW2XA3O6HH/桌面/共享文件/Papers/Prey&Predator/graphics/fig3__3.pdf}%
\caption{Stationary probability distributions of the preys under different
correlations. The parameters are $M_{1}=0.05,$ $M_{2}=0.1$. Inset: Variances
of SPD against the correlations, $\protect\lambda $.}
\end{center}\end{figure}%
\fi
\ifcase\msipdfoutput
\FRAME{ftbpFU}{3.3719in}{2.6965in}{0pt}{\Qcb{Stationary probability
distributions of the predator under different correlations. The parameters
are $M_{1}=0.05,$ $M_{2}=0.1$. Inset: Variances of SPD against the
correlations, $\lambda $.}}{}{fig4.eps}{%
\special{language "Scientific Word";type "GRAPHIC";maintain-aspect-ratio
TRUE;display "USEDEF";valid_file "F";width 3.3719in;height 2.6965in;depth
0pt;original-width 4.0802in;original-height 3.2552in;cropleft "0";croptop
"1";cropright "1";cropbottom "0";filename '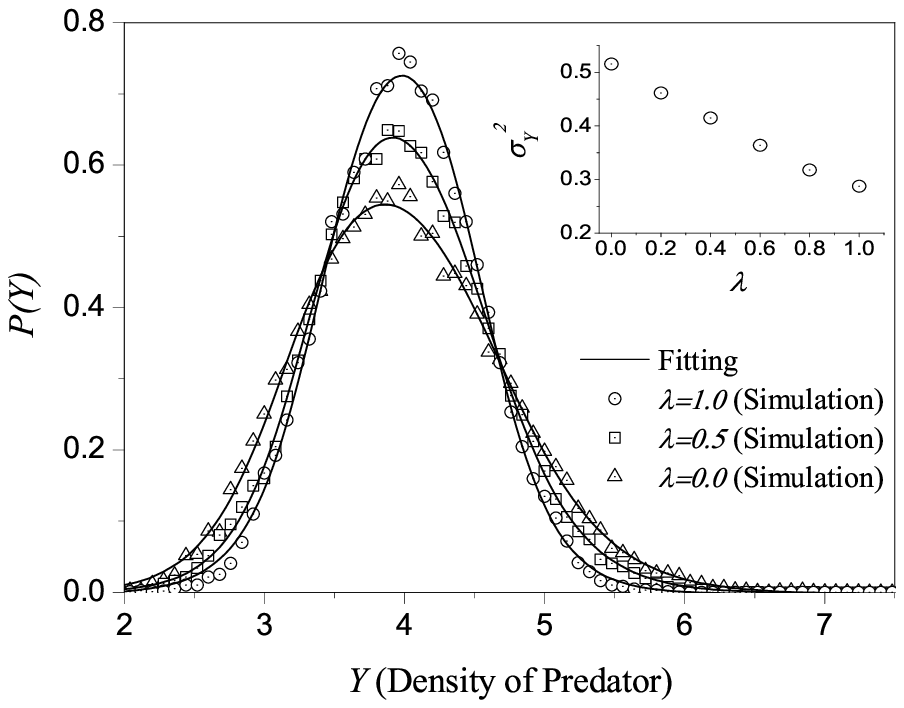';file-properties
"XNPEU";}}%
\else
\begin{figure}[ptb]\begin{center}
\includegraphics[
natheight=3.2552in, natwidth=4.0802in, height=2.6965in, width=3.3719in]
{F:/Documents and Settings/Shao.ZSU-IPW2XA3O6HH/桌面/共享文件/Papers/Prey&Predator/graphics/fig4__4.pdf}%
\caption{Stationary probability distributions of the predator under
different correlations. The parameters are $M_{1}=0.05,$ $M_{2}=0.1$. Inset:
Variances of SPD against the correlations, $\protect\lambda $.}
\end{center}\end{figure}%
\fi

In summary, the correlation or collaboration, which exists in numerous
dynamical systems, can cause interesting responses of the prey-predator
ecosystems to noises. Due to the support of their growth and resource
limitation, the preys undergo perfect endurance to the external noises.
Conversely, the predators are not only sensitive to the noise intensity, but
also impressible to the noise correlation. Strong correlated noises can lead
the stationary probability distributions of the predators to stability. Our
model is expected to support a fact: the complexities of biosystems
originate from their collaborations and correlations.

This work was partially supported by the National Natural Science Foundation
(Grant No. 60471023) and the Natural Science Foundation of Guangdong
Province (Grant No. 031554), P. R. China.

\end{document}